\documentclass[journal]{new-aiaa}
\usepackage[utf8]{inputenc}
\usepackage{textcomp}
\usepackage{graphicx}
\usepackage{amsmath}
\usepackage[version=4]{mhchem}
\usepackage{siunitx}
\usepackage{longtable,tabularx}
\title{Experimental Investigation of Heat Transfer in Axisymmetric Open Cavities at Hypersonic Speeds}

\author{Talluri Vamsi Krishna \footnote{Post-Doctoral Fellow, Technion Wind Tunnels Complex, Faculty of Aerospace Engineering, Email: sri.vamsi1432@gmail.com} and Jacob Cohen \footnote{Professor Emeritus, Technion Wind Tunnels Complex,Faculty of Aerospace Engineering, Email: aerycyc@technion.ac.il}}
\affil{Technion-Israel Institute of Technology, Haifa-3200003, Israel.}
\author{S.K. Karthick \footnote{Assistant Professor, Mechanical and Aerospace Engineering, Email: skkarthick@mae.iith.ac.in }}
\affil{Indian Institute of Technology Hyderabad, Kandi-502285, Telangana, India}
\author{Soumya R. Nanda\footnote{ Assistant Professor, Department of Aerospace Engineering, Email: srnanda@iitj.ac.in }}
\affil{Indian Institute of Technology Jodhpur, NH 62 Nagaur Road, Karwar-342030, Rajasthan, India}

\begin{document}

\maketitle

\begin{abstract}
An experimental investigation was conducted to characterize the heat-transfer behavior of an axisymmetric cavity mounted on a cone with a semi-apex angle of $(48^ \circ)$ under Mach 6 freestream conditions in a Ludwieg tunnel. The effects of cavity geometry, represented by the length-to-depth ratio $(L/D = 0.5, 1.0, 2.0, 4.0,$ and $6.0)$, rear-wall height-to-depth ratio $(h/D = -0.25, 0,$ and $0.5)$, and Reynolds number ($Re_D = 23000 - 74000$) were systematically investigated. Flow-field characteristics and surface heat-transfer behavior were examined using schlieren and planar laser Rayleigh scattering visualization techniques, together with surface pressure and heat-transfer measurements. The results indicate that increasing the cavity length-to-depth ratio leads to a reduction in the downstream Stanton number $((St)_d)$. For the longest cavity $L/D = 6$, however, the downstream Stanton number was found to exceed the upstream value $((St)_u)$ indicating enhanced heat transfer in the region downstream of the cavity.
\end{abstract}

\section*{Nomenclature}

{\renewcommand\arraystretch{1.0}
\noindent\begin{longtable*}{@{}l @{\quad=\quad} l@{}}

$c$        & Specific heat capacity (J/kg$\cdot$K) \\

$D$        & Cavity depth (m) \\

$h$        & Excess height of the trailing edge of the cavity (m) \\

$k$        & Thermal conductivity (W/m$\cdot$K) \\

$L$        & Length of the cavity (m) \\

$M$        & Mach number \\

$\mu$      & Dynamic viscosity (kg/m$\cdot$s) \\

$p$        & Instantaneous static pressure (Pa) \\

$Sp$        & Pressure sensor \\

$Re$       & Reynolds number per m ($\rho u/\mu$, 1/m) \\

$S$        & Streamwise distance along the shear layer (m) \\

$St$       & Stanton number \\

$t$        & Time (ms) \\

$T$        & Instantaneous static temperature (K) \\

$S_T$        & CSJT \\

$u$        & Velocity (m/s) \\

$\rho$     & Density (kg/m$^3$) \\

$\theta$   & Semi-apex angle ($^\circ$)\\

$ FFT $      &  Fast Fourier Transform\\
$ K-H $      & Kelvin-Helmholtz \\
$ PLRS $     & Planar Laser Rayleigh Scattering \\
\multicolumn{2}{@{}l}{Subscripts}\\
 
 $_0$        & Total (stagnation) quantity \\

 $_2$        & Quantity at the cavity entrance \\

 $_\infty$   &  Freestream quantity \\

 $\beta$    & Effective thermal product \\

 $_r$        & Rear face \\

 $_D$        & Based on cavity depth  \\

 $_{fc}$       & Floor center\\

\end{longtable*}}

\addtocounter{table}{-1}

\section{Introduction}
\lettrine{C}{avities} exposed to supersonic and hypersonic flows have been extensively investigated because they occur in various aerospace applications. These cavities may arise intentionally, such as in weapon bays\cite{Loupy}, landing gear compartments\cite{Guo}, and scramjet combustors\cite{Ben_Yakar,CAO2021111562}, or unintentionally due to manufacturing defects, sensor installations\cite{GORBUSHIN2023114095}, gaps in thermal protection systems\cite{petley1984surface}, and differential material ablation. This study examines hypersonic flow over an axisymmetric cavity configuration relevant to applications ranging from thermal protection to scramjet combustors. In small-scale cavities, the separated shear layer can provide thermal shielding by limiting heat transfer to embedded sensors\cite{chapman1958investigation}, whereas large-scale cavities integrated within scramjet combustors are commonly used to improve fuel-air mixing\cite{Gruber2004} and may also function as trapped-vortex combustors \cite{Sturgess1997}and scramjet flame holders \cite{liu2019cavity}. In hypersonic laminar flows, open cavities with suitable $L/D$ ratios ($L/D < 7$) can reduce the heat transfer rate by nearly 50\% \cite{Creighton2007}. With flow reattachment near the cavity trailing edge, a localized hotspot\cite{Nicoll1964} is generated, and the associated heat load increases substantially when the incoming shear layer transitions from laminar to turbulent\cite{Everhart2010}, which can adversely affect the structural integrity and operational life of aerospace vehicles. Therefore, characterizing the shear-layer state, its dynamics, and its interaction with the cavity trailing edge is essential for understanding the mechanisms governing flow unsteadiness and localized aerodynamic heating. The flow over a cavity is inherently complex, involving interactions among the separated shear layer, shock waves, boundary layers, and recirculating flow structures. Several review studies \cite{LAWSON2011186, Creighton2007, Rockwell1978} provide comprehensive discussions on cavity-flow physics and classify the flow behavior based on geometrical parameters: length-to-depth ratio ($L/D$), length-to-width ratio ($L/W$), cavity geometry (rear-wall shapes), and flow parameters such as the depth-to-incoming momentum thickness ratio ($D/\theta_0$), Mach number, and Reynolds number.

The cavity flow field is governed by several coupled mechanisms, including shear-layer instabilities, convection of Kelvin-Helmholtz vortices, acoustic-wave propagation, and unsteady shock motion \cite{Creighton2007}. These flow characteristics strongly depend on the cavity geometry, particularly the aspect ratio ($L/D$). Based on this parameter, cavity flows are generally classified as open$(1 \leq L/D\leq 10)$, transitional $(10 \leq L/D\leq 14)$, or closed cavities $(L/D> 14)$ \cite{guo2019}. Rossiter \cite{rossiter1964} proposed a semi-empirical relation to predict the dominant oscillation frequencies of rectangular cavities as a function of the incoming Mach number and cavity geometry. Later Heller et al.\cite{Heller1971} has modified this formula as,

\begin{equation}
\label{eq:rossiter}
     \left[\frac{fD}{u}\right] = \frac{(n-\alpha)(D/L)}{M\left(1+\left[(\gamma-1)/2\right]M^2\right)^{-1/2}+1/\kappa},
\end{equation}

Here, $M$ denotes the Mach number of the incoming flow immediately upstream of the cavity, while $n = [1, 2, 3, 4, \ldots]$ represents the corresponding mode number. The empirical parameters $\alpha = 0.25$ and $\kappa = 0.57$ correspond to the phase lag of the feedback loop and the velocity ratio of the K-H convection to the freestream, respectively\citep{murray2001}. In the Rossiter-mode framework, the lower-order modes are generally associated with large-scale motions of the separated shear layer. In particular, the fundamental mode ($n=1$) is commonly linked to the global flapping of the shear layer, which promotes periodic exchange of fluid between the cavity and the external stream \cite{rezende2024}. As the mode number increases, this large-scale motion progressively interacts with the shear-layer convective instability. Consequently, the second and third modes can involve both global flapping and the development of Kelvin–Helmholtz (K–H) vortices, with coherent vortical structures forming and being convected downstream along the shear layer\cite{mathias2021effect}. For higher Rossiter modes ($n \geq 4$), the instability dynamics shift toward progressively smaller streamwise scales, resulting in shorter-wavelength K–H structures and correspondingly higher characteristic frequencies. The spatial coherence of these vortices, however, tends to deteriorate as the freestream Mach number increases\cite{murray2001}. At high supersonic conditions, the feedback process becomes more complex than in the subsonic cavity owing to the relative propagation speeds of large-scale convecting structures, the external freestream, and the flow within the cavity\cite{bres2008,sandham1991}. Furthermore, increasing Mach number promotes a transition toward increasingly three-dimensional dynamics, with 3D instability modes becoming more prominent relative to their predominantly two-dimensional counterparts\cite{islam2024,doshi2022,garrido2014}. These observations indicate that compressibility does not merely alter the strength of the classical Rossiter modes but can also modify the underlying instability mechanisms and the dimensionality of the resulting cavity-flow structures.

Heat transfer in the vicinity of the cavity depends on the length-to-depth $(L/D)$ and width-to-depth $(W/D)$ ratios. Higher $L/D$ values and wider cavities promote earlier boundary-layer transition, resulting in increased aerodynamic loads and higher heat transfer rates\cite{Liechty2006}. Heat transfer studies over open and closed cavities at hypersonic speeds showed that a closed cavity has significantly higher heat transfer on the cavity floor due to the reattachment region, while an open cavity has significantly less heat transfer on the cavity floor than a closed cavity\cite{NESTLER}. The extensive investigations of cavity-induced aerodynamic heating under supersonic and hypersonic flow conditions focus on the effects of pressure gradients, cavity geometry, and incoming boundary-layer characteristics\cite{Everhart2006,Everhart2007shallow, Everhart2009, Everhart2010,Everhart2011}. The experimental results from these studies indicated that pressure gradients could increase the average cavity-floor heating by up to 50\% in closed cavities, while their influence on open-cavity heating was relatively limited under laminar boundary-layer conditions at the cavity entrance and exit \cite{Everhart2006}. Further investigations examined cavity heating associated with impact-induced damage to the Space Shuttle's thermal protection system, demonstrating the importance of cavity length and depth, and their relationship to the incoming boundary-layer thickness, in determining local heating augmentation. These studies provided an experimental basis for characterizing cavity-induced heating and developing predictive methods for estimating localized thermal loads in hypersonic vehicles. Building on these experimental investigations, Everhart \cite{Everhart2009} developed empirical correlations to predict the average cavity-floor heating, peak heating near the downstream cavity end wall, and the spatial decay of heating downstream of the cavity. The correlations accounted for cavity geometry and local flow conditions and were evaluated for both open and closed cavities, as well as for non-baseline configurations involving non-rectangular geometries and pressure gradients.

Although cavity flows have been extensively investigated, most existing studies have focused on two-dimensional rectangular cavities, with comparatively little attention to axisymmetric configurations. In particular, the effects of flow and geometric parameters on the local heat transfer characteristics arising from shock–shear-layer interactions in hypersonic axisymmetric cavity flows remain insufficiently understood. This study addresses this issue by investigating the heat transfer characteristics of a cone-mounted axisymmetric open cavity under hypersonic flow conditions. Specifically, we examine the effects of Reynolds number $(Re_D)$, cavity aspect ratio $(L/D)$, and rear-wall height-to-depth ratio $(h/D)$ over a broad range of operating and geometric conditions to elucidate their influence on shock–shear-layer interactions and the resulting local heat transfer distributions.

\section{Experimental Methodology}

 \subsection{Experimental facility and Model details}
 
\begin{figure}[h!]
\centering
\includegraphics[width=1\textwidth]{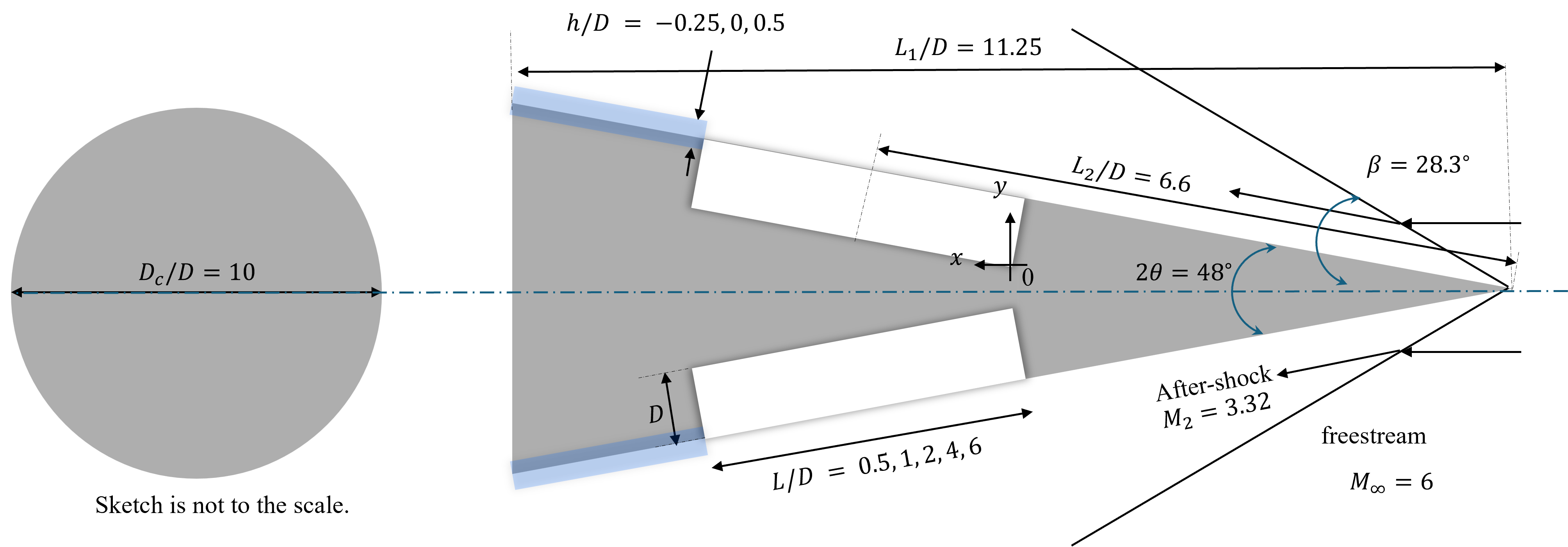}
\caption{Schematic of the experimental model used in the present study.}
\label{fig:model}
\end{figure}
In the present investigation, all the experiments are conducted using the Ludwieg tunnel available at the Technion Wind Tunnel Complex. This tunnel can produce a freestream Mach number of 6 with Reynolds numbers ranging from $Re_\infty=4.7-16.7\times 10^6 (m^{-1})$, obtained by varying the driver tube fill pressure. Uniform flow is obtained in the test section for a duration of 12.5 ms. More details on the experimental facility, flow quality, and available instrumentation are given in the authors' earlier work \cite{Karthick_2023}. 

\begin{figure}[h!]
\centering
\includegraphics[width=0.7\textwidth]{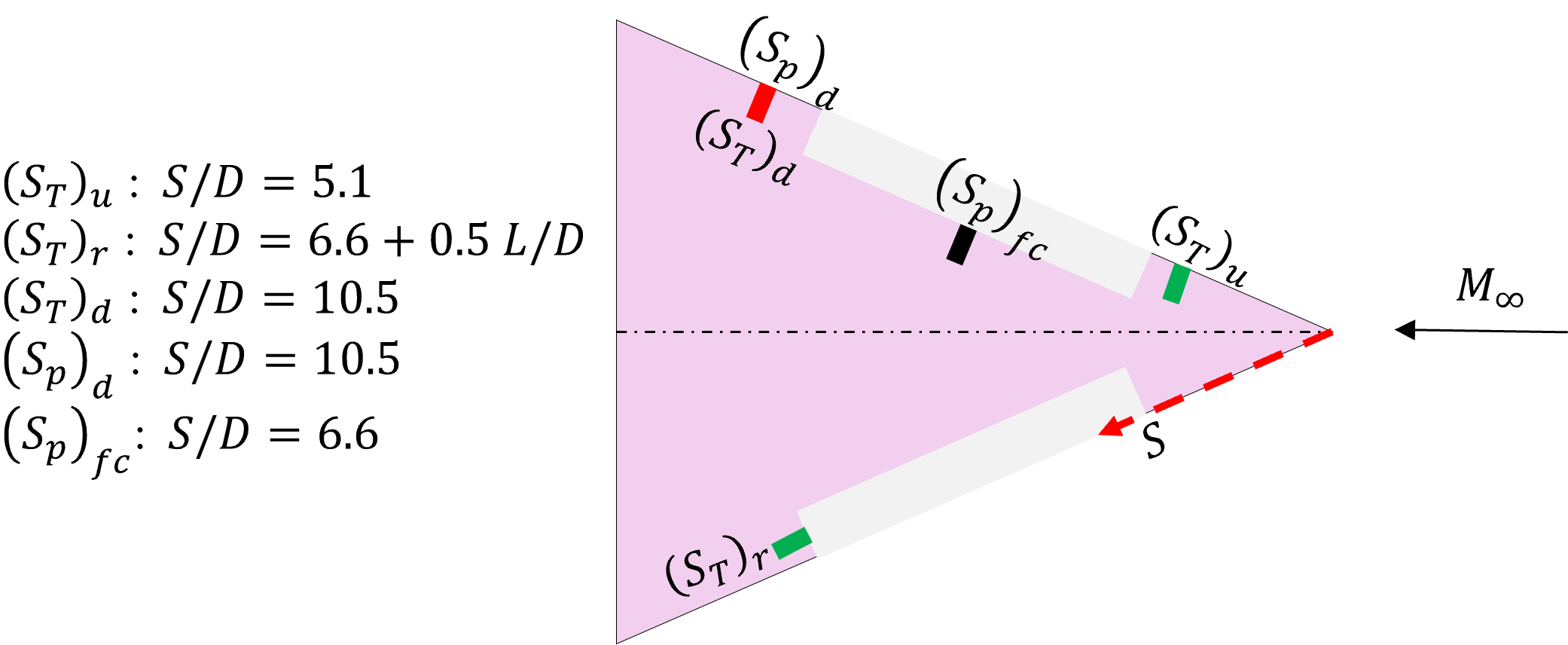}
\caption{Schematic showing the locations of the pressure transducers and thermocouples mounted on the cavity model. }
\label{fig:Sensor}
\end{figure}

Experiments are carried out using an axisymmetric cavity configuration mounted on a cone (as shown in Figure \ref{fig:model}) with an apex angle of $2\theta = 48^\circ$. The  cavity length ($L$) to depth ($D$) ratio is varied in the range $[L/D]= [0.5,1.0,2.0,4.0,6.0]$. Different axisymmetric cavity models are constructed with a constant depth of $D = 4$ mm.  For all the cavity configurations, the center of the cavity is placed at a distance of $6.6D$ from the cone tip. In addition, the excess rear wall height $ h/D$ is varied over $ [-0.25, 0, 0.5]$ for $L/D=1.0$ and $6.0$, effectively resulting in rear face depths of 3, 4, and 6 mm respectively keeping the front wall depth of 4 mm. During the experimental campaign, the $Re_D$ (see Table \ref{table:Freestream conditions}) is changed by altering the stagnation pressure ($p_{01}$) of the incoming flow. The flow properties in the freestream and ahead of the cavity for the extreme cases available for the current study are shown in Table \ref{table:Freestream conditions}. The present investigation focuses primarily on quantifying the mean surface pressure and heat loading across different zones of the cone-cavity configurations. The pressure sensors (Kulite\textsuperscript{\textregistered} XCQ-080-100A) are mounted on the cavity floor center ($(S_p)_{fc}$), and at a location downstream the cavity, $(S_p)_d$ ($S/D = 10.5$) as shown in Figure \ref{fig:Sensor}. To estimate the local heat flux, E-type coaxial surface junction thermocouples \cite{heufer2008} (CSJT) (RWTH-Aachen HGD\textsuperscript{\textregistered} KLP and MI-3.6) are installed at the rear face center ($(S_T)_r$) and downstream of the cavity ($(S_T)_d$), for all the cases under consideration.  Due to space limitations, only for $L/D = 0.5, 1.0,$ and 2.0, a CSJT is mounted upstream of the cavity ($(S_T)_u$) at $S/D = 5.1$ to evaluate the local heat flux from the attached incoming flow to the cavity (See Figure \ref{fig:Sensor}). To ensure repeatability, each experiment is repeated three times for each configuration. CSJT and pressure sensor signals are recorded using National Instruments data acquisition cards at a sampling rate of 100 kHz and 800 kHz, respectively. Flow visualization experiments are performed using a Z-type Schlieren arrangement incorporating parabolic and plane mirrors, each with a diameter of 150 mm and a focal length of 1.5 m. To accurately capture the spatio-temporal characteristics of the flow, images are acquired at a high spatial resolution of 1280 × 800 pixels with a recording rate of 2.2 kHz and an exposure time of 2 $\mu s$ using a Phantom v211 high-speed camera. Planar Laser Rayleigh Scattering (PLRS) measurements were conducted for selected test cases to characterize the flow dynamics of the cavity. PLRS snapshots are recorded at a spatial resolution of 128 × 96 pixels corresponding to an acquisition frequency of 100 kHz.

 \begin{table}[h!]
\caption{The flow properties in the freestream and cavity upstream for the extreme cases under discussion.}
\centering
\begin{tabular}{lcc}
 \hline
\textbf{Parameters} & \textbf{Freestream ($\infty$)} & \textbf{Cavity upstream ($2$)}\\\hline
 Total pressure ($p_{01} \times 10^5$, Pa) & 2.31 - 8.13 & 0.86 - 3.04\\
Total enthalpy in the driver ($h_{01}$, MJ/kg) & 0.3 & 0.3\\
Pressure ($p$, Pa) & 152.64 - 514.92 & 1472 - 4785\\
Temperature ($T$, K) & 36.58 & 91.6\\
Density ($\rho$, kg/m$^3$) & 0.015 - 0.049 & 0.056 - 0.182\\
Kinematic viscosity ($\nu\times 10^{-5}$, m$^2$/s) & 14.7 - 4.4 & 11.2 - 3.4\\
Velocity ($u$, m/s) & 727.46 & 636.93\\
Mach number ($M$) & 6.0 & 3.32\\
Reynolds number ($Re_D \times 10^4$, $D$: cavity depth) & 1.9 - 6.7 & 2.3 - 7.4\\
\hline
\end{tabular}
\label{table:Freestream conditions}
\end{table}

 \subsection{Thermocouple data processing}

Heat transfer measurements were conducted using an E-type CSJT, having a sensitivity of 59.8 $\mu V/K$. The sensor output was pre-amplified by a factor of 2000 using an in-house-fabricated amplifier before being supplied to the NI data acquisition system. The temperature rise was measured during the test, and Equation \ref{eq:HF_eqn} was used to determine the surface heat flux ($\dot{q}_s$). The heat flux was estimated under the assumption  of the sensor being a semi-infinite body undergoing one-dimensional heat conduction. The detailed analytical procedure used for this estimation is provided by Taler \cite{TALER19963733}.

\begin{equation}
\label{eq:HF_eqn}
\dot{q_s} = \frac{2\beta}{\sqrt{\pi}} \sum_{i=1}^{n} \frac{T(t_i) - T(t_{i-1})}{{{(t_n-t_i)}^{1/2}} + {{(t_n-t_{i-1})}^{1/2}}},\; \text{and}\; \beta =\sqrt{\rho c k},
\end{equation}

where $\rho$ is the density, $c$ is the specific heat capacity, $k$ is the thermal conductivity of the substrate material, and $\beta$  $($8500 $J/m^2KS^{0.5}$) is defined as the `effective thermal product (TP) of the substrate material.

\begin{equation}
\label{eq:stanton}
St = \left[\frac{\dot{q_s}}{\rho_2 u_2 (h_{2}-h_w)}\right], \; \text{where, } h_{2}=C_p\times T_{2} \; \text{and} \; \ h_{w}=C_p\times T_{w}.
\end{equation}

The local heat flux is estimated from the rise in temperature signal recorded using CSJT at three locations as shown in Figure \ref{fig:Sensor}, and the variation of the heat flux with $L/D$, $Re_D$ and $h/D$ is then assessed. The non-dimensional heat flux is represented in terms of the Stanton number ($St$) and calculated using Equation \ref{eq:stanton} where $\rho_2$, $u_2$, and $T_{2}$ corresponds to the cavity upstream density, velocity and static temperature, respectively as shown in Table \ref{table:Freestream conditions}; whereas, $T_{w}$ represents the initial temperature of the sensing junction of CSJT taken as $300$ K.

\begin{figure}[h!]
\centering
\includegraphics[width=1\textwidth]{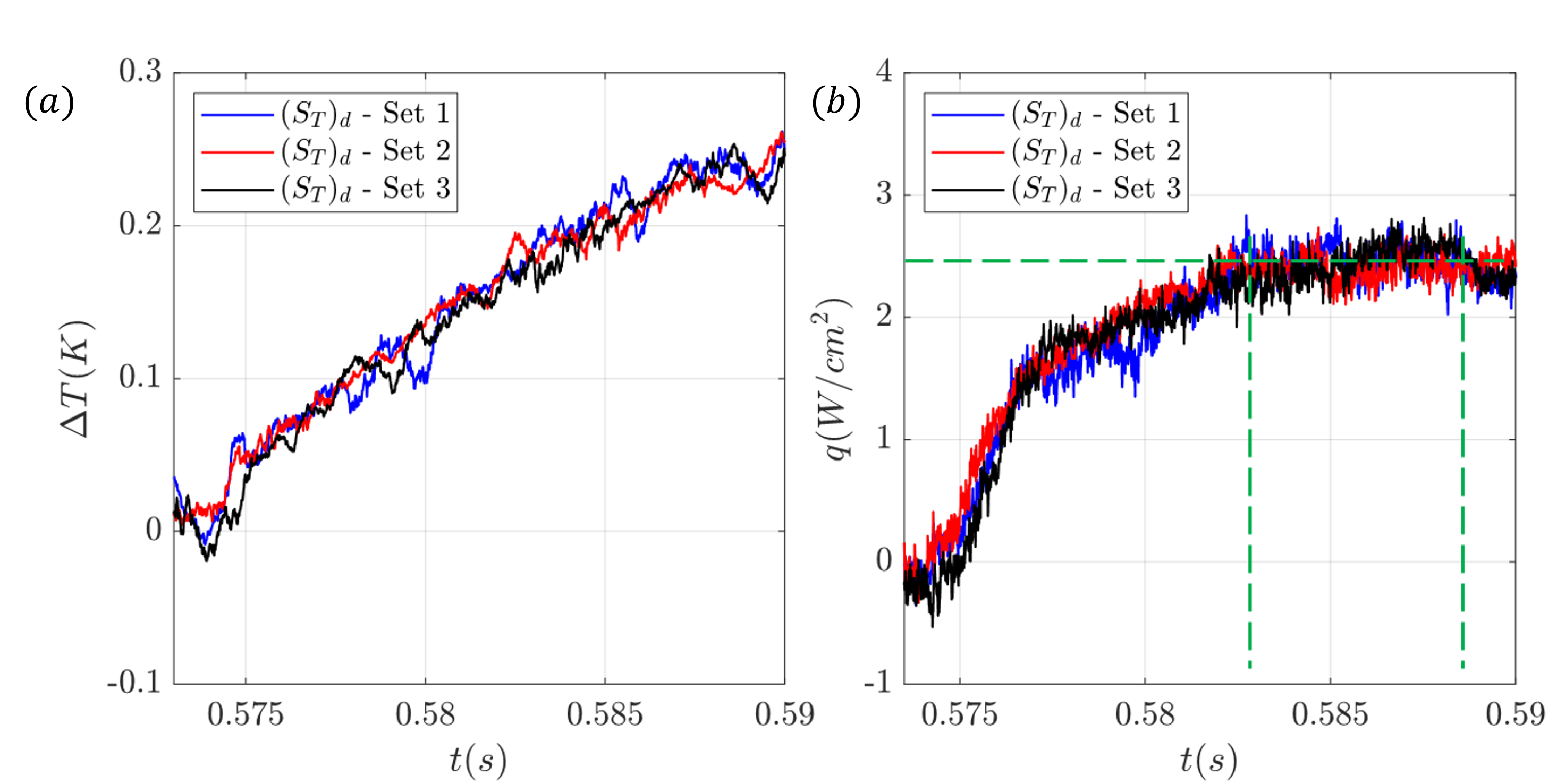}
\caption{Plot showing (a) rise in temperature and (b) heat flux signal for three runs obtained from the CSJT mounted at the cavity downstream location: $(S_T)_d$ for the case of $[L/D, h/D, Re_D] = [6.0, 0.5, 74000]$. The green horizontal line demarcates the steady heat flux value, and the time gap between the two vertical green lines indicates the time duration during which heat flux is considered to be constant.}
    \label{fig:Repeatability HF}
  \end{figure}

Typical rise in temperature during the test time and its repeatability across three runs are presented in Figure \ref{fig:Repeatability HF}(a). The local heat flux ($\dot{q_s}$) is estimated using Equation \ref{eq:HF_eqn}, and the repeatability of the measured responses is illustrated in Figure \ref{fig:Repeatability HF}(b). The calculated heat flux values are consistent across all three runs, providing confidence in the measured heat transfer during the test time. The steady heat flux values marked by the green color dashed horizontal line in Figure \ref{fig:Repeatability HF}b are used for calculating $St$ and for further comparison.

\begin{figure}[h!]
    \centering
    \includegraphics[width=0.5\linewidth]{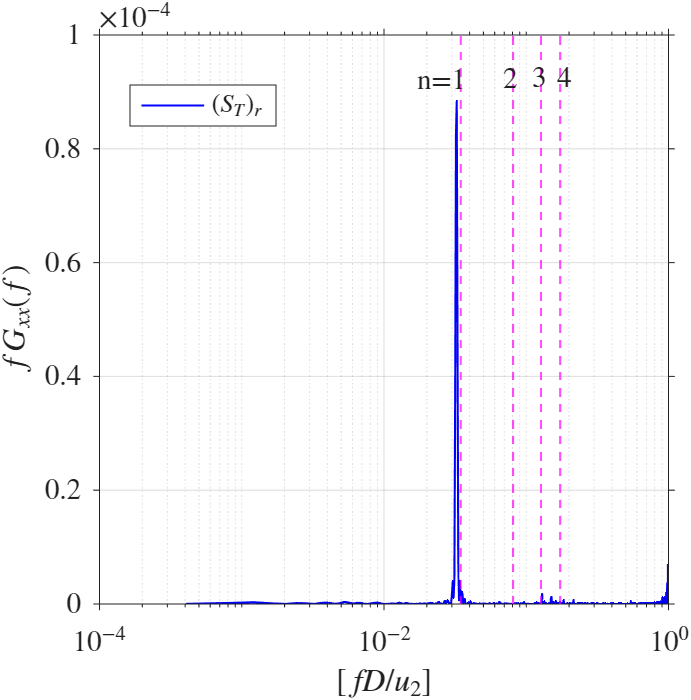}
    \caption{Frequency spectra obtained from the CSJT sensor on the rear wall at $L/D = 6.0, h/D = 0.5$, and $Re_D = 74000$. Dashed lines represent the Rossiter modes.}
    \label{fig:freq}
\end{figure}

The temporal response capability of the thermocouple was assessed by comparing its spectral characteristics with the Rossiter modes using equation \ref{eq:rossiter}. As shown in Figure \ref{fig:freq}, both the CSJT sensor at cavity rear-wall $((S_T)_r)$ exhibit a pronounced peak at the frequency corresponding to the first Rossiter mode, indicating that the dominant characteristic frequency is consistently captured by the currently employed CSJTs. For this configuration, Nanda et al.\cite{nanda2026} also reported the same dominant mode at the same Reynolds number based on measurements obtained using pressure sensors and planar laser Rayleigh scattering (PLRS). The presence of the same dominant frequency in the thermocouple signal demonstrates that its temporal response is sufficient to resolve the primary unsteady fluctuations associated with the cavity flow.

\section{Results and Discussion}

\subsection{ Aspect $(L/D)$ Ratio Effect}
\subsubsection{ Schlieren Flow Visualization }

\begin{figure*}[]
\centering
\includegraphics[width=0.8\textwidth]{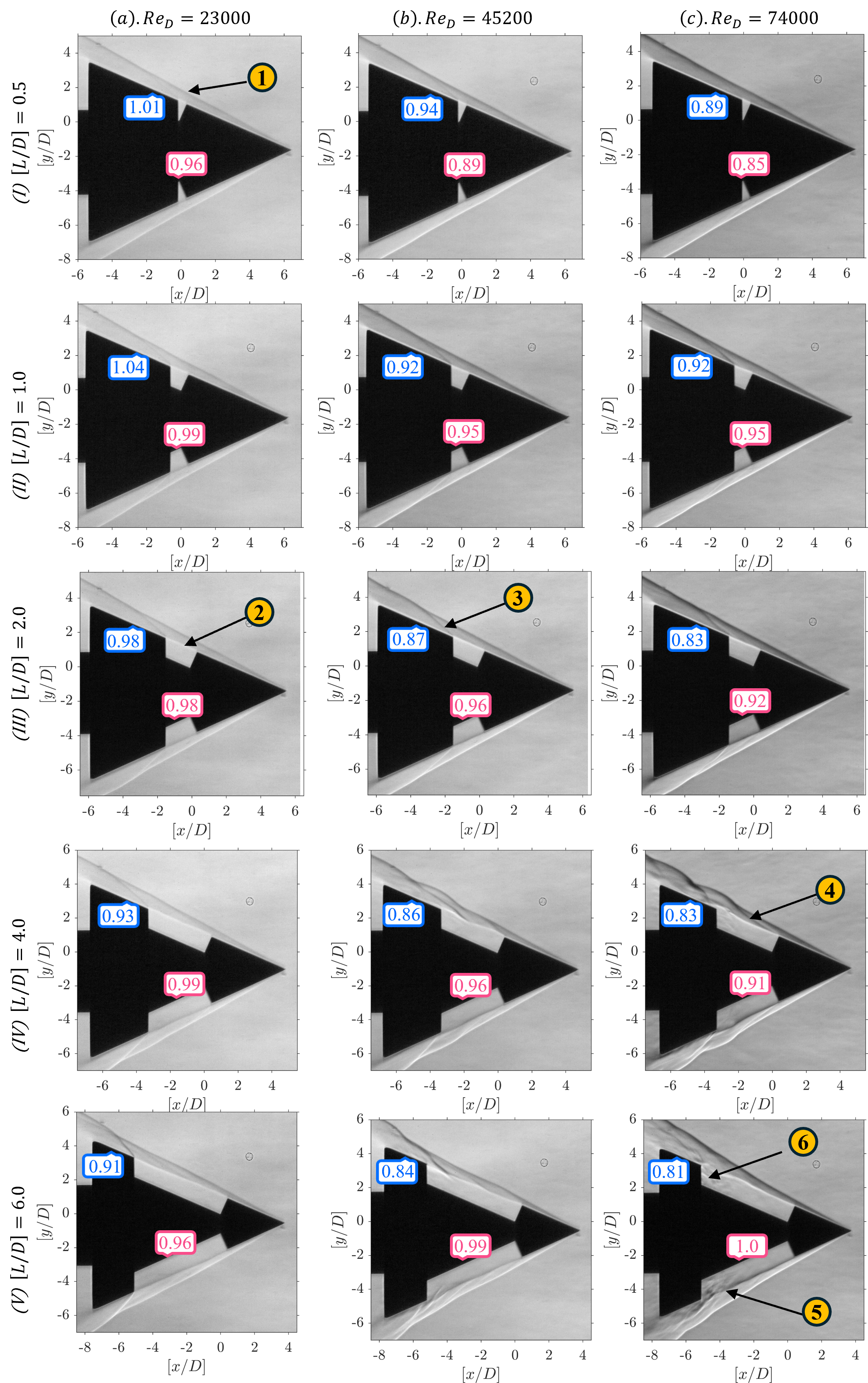}
\caption{Schlieren flow visualization for cavity configuration corresponding to $L/D=[0.5,1.0, 2.0,4.0,6.0]$ and $Re_D=[23000,45200,74000]$. 1. leading edge conical shock, 2. separated shear layer, 3. reattachment shock at the trailing edge, 4. shock wave from the undulating shear layer, 5. shear layer turbulent breakdown, and 6. upstream propagation of acoustic waves.   }
\label{fig:Sch_lbyd}
\end{figure*}

Initially, schlieren flow visualization experiments are conducted to investigate cavity flow behavior at different $L/D$ ratios with $h/D = 0$. Instantaneous schlieren images for $L/D = [0.5, 1.0, 2.0, 4.0, 6.0]$ at Reynolds numbers $Re_D  = [23000, 54200, 74000]$ are depicted in Figure \ref{fig:Sch_lbyd}. These Schlieren snapshots reveal important flow features, including the leading-edge shock, shear layer, and reattachment shock as identified in Figure \ref{fig:Sch_lbyd}. In addition to the schlieren flow visualization, the normalized mean pressure ($p/p_2$) was measured at the cavity center ($(Sp)_{fc}$) and downstream ($(Sp)_d$), with the sensor locations marked at their respective positions in the images. 

Schlieren flow visualization in Figure \ref{fig:Sch_lbyd} indicates that, for most cavity configurations, the shear layer separates from the cavity front wall and reattaches at the rear wall, confirming open-cavity flow behavior. For $L/D$ ratios of $0.5$, $1.0$, and $2.0$ (Figure \ref{fig:Sch_lbyd} I-III), the separated shear layer exhibited no noticeable vortex roll-up at all the Reynolds numbers investigated. A similar trend is also observed for the larger cavities ($L/D = 4.0$ and $6.0$) at the lowest Reynolds number, $Re_D  = 23000$, indicating the presence of a predominantly laminar shear layer. However, for $L/D$ ratios of $4.0$ and $6.0$, increasing the Reynolds number leads to the development of coherent vortical structures near the rear-wall impingement region due to Kelvin–Helmholtz instability (see Figure \ref{fig:Sch_lbyd}-IV(b) and V(b)). This effect becomes more pronounced at higher Reynolds numbers, as the feedback mechanism is intensified either when $Re_D$ or the $L/D$ ratio is increased. In addition, both the shock motion and the separated shear-layer oscillations intensified as the Reynolds number is increased, leading to a substantial strengthening of the reattachment shock. At the highest Reynolds number, $Re_D  = 74000$, for $L/D = 6.0$, acoustic waves generated at the trailing edge and propagating upstream inside the cavity are clearly visible. These acoustic disturbances enhance the growth of Kelvin–Helmholtz vortices along the separated shear layer, causing them to amplify in the streamwise direction and eventually resulting in a plausible transition to turbulence. The disappearance of shear-layer waviness marks the end of the laminar region and the onset of turbulent flow at $Re_D  = 74000$ for $L/D = 6.0$ (see Figure \ref{fig:Sch_lbyd}-V(c)). As reported earlier \citep{nanda2026}, axisymmetric cavities exhibit an upstream shift in the shear layer perturbation location with increasing $Re_D$ along with possible azimuthal structure within the cavity can be probable reason for the laminar to turbulent transition in axisymmetric cavities.

The normalized static pressure ($\zeta =p/p_2$) at the cavity floor center ($(Sp)_{fc}$) for $L/D=0.5$ depicts a gradual decrement trend with an increase in $Re_D$, which essentially indicates a lower rate of pressure recovery. For $L/D=0.5$, the separated shear layer has very minimal streamwise distance to develop, thereby limiting the interaction of cavity recirculation and overlying shear layer. Moreover, comparison of the schlieren snapshots in Figures \ref{fig:Sch_lbyd}-I(a) and \ref{fig:Sch_lbyd}-I(c) reveals similar flow features without the presence of any additional structures. Therefore, with an increase in $Re_D$, no additional mechanism of pressure recovery is evident, resulting in the reduction of $\zeta$. As the $L/D$ increases, the longer streamwise length allows the shear layer to evolve and interact with the cavity recirculation. The enhanced interaction promotes pressure recovery within the cavity through a feedback loop and shear layer unsteadiness. Consequently, the reduction in the rate of decrement in $\zeta$ at the location of $(Sp)_{fc}$ becomes less pronounced as $Re_D$ increases for a particular $L/D$. However, for $L/D=6.0$, owing to the transitional shear layer behaviour at higher $Re_D$ cases, it deviates from the aforementioned trend and exhibits $\zeta \approx 1$ (see Figure \ref{fig:Sch_lbyd}-V).

The downstream pressure distribution is governed by the combined effects of the cavity $L/D$ ratio and $Re_D$. At smaller $L/D$ ratios and $Re_D$ combinations, the separated shear layer remains nearly attached to the external flow and spans the cavity with minimal penetration. Under these conditions, the incoming boundary layer experiences only a weak influence from the cavity geometry, resulting in downstream pressure values that remain equal to the incoming flow pressure (see Figure \ref{fig:Sch_lbyd} I-III(a)). As the cavity $L/D$ ratio increases, the separated shear layer penetrates further into the cavity, and the flow becomes increasingly sensitive to the cavity geometry. For $L/D\geq2$, especially at lower Reynolds numbers ($Re_D=23000$), the shear layer reattaches at the trailing edge, giving rise to a reattachment shock that contributes to pressure recovery downstream. However, increasing the Reynolds number ($Re_D=45200$ and $74000$) alters the shear-layer dynamics by promoting the formation and downstream convection of K-H vortices (see Figure \ref{fig:Sch_lbyd} -V(b)-(c)). These vortices interact strongly with the cavity trailing edge, strengthening the reattachment shock while simultaneously increasing its unsteadiness. The interaction between the vortices and the reattachment shock spreads the reattachment process over a larger streamwise distance, making the flow turning more gradual and distributed. As a consequence, the downstream pressure recovery becomes less effective for cavities with $L/D=2.0, 4.0,$ and $6.0$, leading to lower mean pressure values as the Reynolds number increases (see Figure \ref{fig:Sch_lbyd} III-V(c)). In contrast, at lower $Re_D$s, the weaker density gradients near the trailing edge indicate smoother flow turning and a more stable reattachment process, resulting in improved pressure recovery and comparable pressure levels between the cavity floor and the downstream region (see Figure \ref{fig:Sch_lbyd}(a)). Overall, regardless of the cavity $L/D$ ratio, the $\zeta$ decreases with increasing Reynolds number as enhanced shear-layer instability and distributed reattachment reduce the effectiveness of pressure recovery downstream of the cavity.

 \subsubsection{PLRS Visualization}
 
 \begin{figure}[hbt!]
\centering
\includegraphics[width=1.0\textwidth]{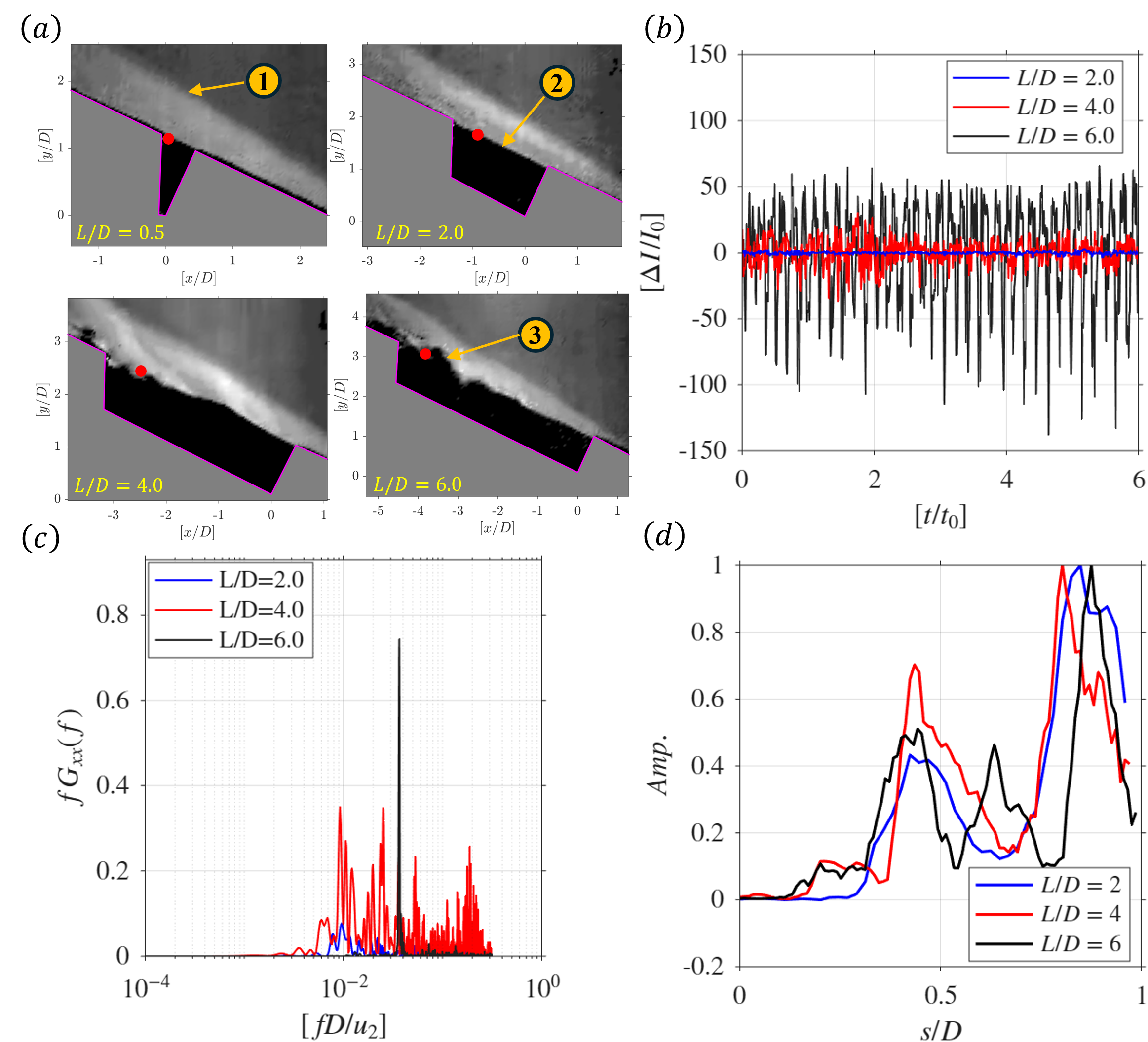}
\caption{ (a) Instantaneous PLRS images obtained for the  $L/D=0.5, 1.0, 2.0$ and $6.0$ at $Re_D=74000.$ (b) Fluctuating light intensity at the point marked in the PLRS snapshots (c) FFT of the shear layer intensity fluctuations (d) Amplitude variation along the shear layer of the cavity for the corresponding dominant frequency. Flow features: (1) leading edge shock (2) laminar shear layer (3)  Kelvin–Helmholtz (K–H) vortices   }
\label{fig:PLRS_FFT}
\end{figure}

 Figure \ref{fig:PLRS_FFT}(a) presents the instantaneous PLRS visualizations for $L/D=0.5, 1.0$, $2.0$, and $6.0$ at $Re_D=74,000$. The PLRS images capture several prominent flow features, including the leading edge shock, the separated shear layer, and Kelvin–Helmholtz (K–H) vortices developing along the shear layer. For $L/D=0.5$, the separated shear layer over the cavity remains predominantly laminar, and a similar behaviour is observed for $L/D=1.0$ and $2.0$. However, as the cavity aspect ratio increases to $L/D=6.0$, the shear layer undergoes a transition from a predominantly laminar state to a more unstable state, with distinct Kelvin–Helmholtz vortices clearly visible in the shear layer. The intensity fluctuations measured in the shear layer at 90 \% of the imaginary line connecting the leading and trailing edge for $L/D=0.5$, $1.0$, and $6.0$ are shown in Figure \ref{fig:PLRS_FFT}(b). For $L/D=0.5$, the intensity fluctuation signal remains nearly flat, indicating a relatively stable shear layer. As the aspect ratio increases to $L/D=1.0$, the fluctuation intensity increases, suggesting the onset of weak unsteadiness within the shear layer. Further increasing the aspect ratio to $L/D=6.0$ results in clear large periodic oscillations of the shear layer. This behaviour is consistent with the flow features observed in the earlier schlieren visualizations (see Figure \ref{fig:Sch_lbyd}). Fast Fourier transform (FFT) analysis was applied to the intensity signals to investigate the dominant frequency characteristics of the cavity flow, as shown in Figure \ref{fig:PLRS_FFT}(c). For $L/D=0.5$ and 1.0, the measured frequency spectra do not exhibit any distinct discrete peaks and are characterized by a relatively broadband and random unsteadiness. In contrast, for $L/D=6.0$ , the spectrum exhibits a clear dominant peak at $fD/u_2 = 0.19$, which corresponds to the third Rossiter mode. Similar discrete unsteadiness is also observed for $L/D=2.0$ and $4.0$ across all the $Re_D$s considered herein.  For the cases where a distinct peak is identified in the frequency spectrum, the amplitude corresponding to the dominant frequency is extracted along the cavity shear layer to examine the associated spatial distribution of the unsteady flow structures. Figure \ref{fig:PLRS_FFT}(d) presents the amplitude distribution along the cavity shear layer for $L/D=2.0$ , 4.0, and 6.0, evaluated at the corresponding dominant frequencies. For $L/D = 2.0$, two distinct amplitude jumps are observed, indicating the presence of two coherent flow structures spanning the entire cavity. As the aspect ratio increases to $L/D = 4.0$ and 6.0, three and four distinct amplitude jumps are observed, respectively. These observations are consistent with the POD analysis reported by Nanda et al. \cite{nanda2026}, which identified the same number of dominant flow structures for the respective cavity configurations.

\subsubsection{ Surface Heat Flux Measurement }
\begin{figure}[hbt!]

\centering
	\includegraphics[width=1.0\textwidth]{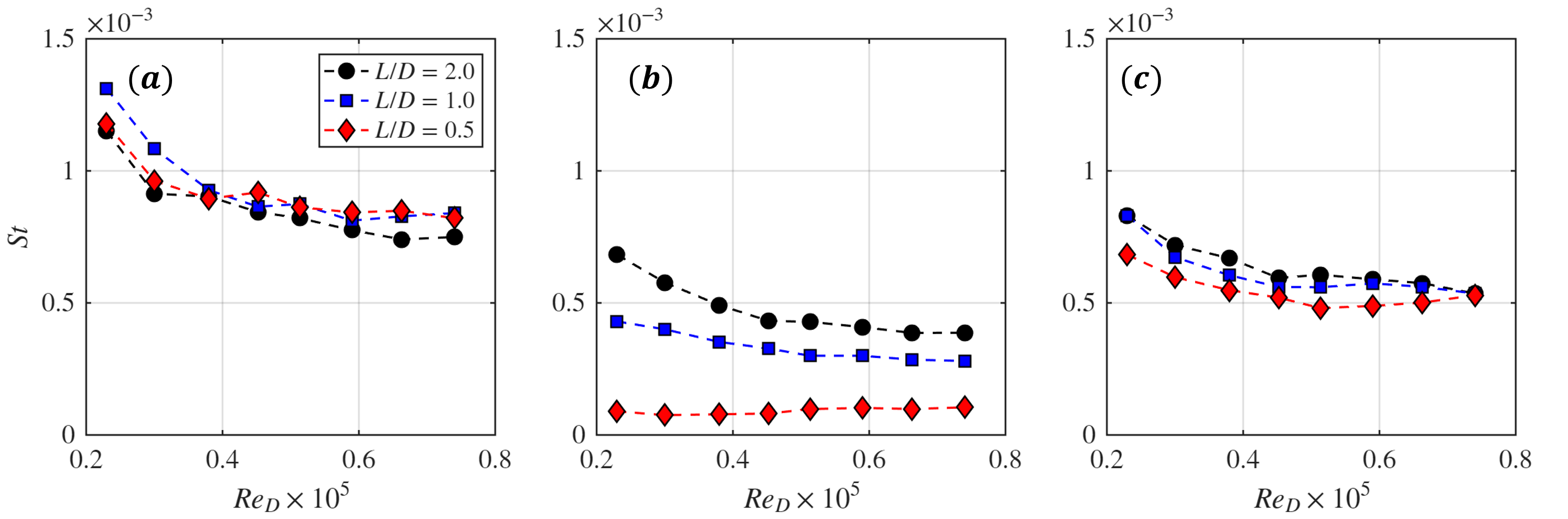}
	\caption{Local Stanton number variation obtained from (a)$(S_T)_{u}$ (b) $(S_T)_{r}$ (c)  $(S_T)_{d}$  (See Figure \ref{fig:Sensor} for sensor details) over $23000\leq Re_D \leq 74000$ for  $L/D$ ratios of 0.5, 1.0 and 2.0.}
    \label{fig:HF_LbyD_effect_0p5t2}
\end{figure}

The local heat flux variation obtained in terms of $St$ from the temperature rise using CSJT on the cavity upstream, rear face, and downstream for $L/D = [0.5, 1.0, 2.0]$, across all the $Re_D=23000-74000$, is shown in Figure \ref{fig:HF_LbyD_effect_0p5t2}. The $St$  variation obtained upstream of the cavity (see Figure \ref{fig:HF_LbyD_effect_0p5t2}(a)) shows a reasonable match in magnitude across all the $L/D$ cases, especially at higher $ Re_D$, thereby indicating an invariant behavior of $St$ with the cavity aspect ratio. This also indicates that the flow upstream of the cavity is unaffected by the cavity's presence. The Stanton number obtained on the cavity rear face is shown in Figure \ref{fig:HF_LbyD_effect_0p5t2} (b). For $L/D = 0.5$, as discussed in the preceding section, the flow within the cavity remains largely isolated from the external flow. This flow isolation leads to a significantly lower $St$ compared with those observed for the other $L/D$ ratios. Moreover,  $St$ exhibits negligible variation with increasing $Re_D$, indicating that the characteristic flow dynamics within the cavity are relatively insensitive to changes in $Re_D$ for $L/D=0.5$. As the $L/D$ ratio increases to $1.0$ and $2.0$, a significant increase in the $St$ is observed with corresponding $Re_D$ and $St$ is decreasing as $Re_D$ increases. This behaviour can be attributed to the larger cavity aspect ratios, which facilitates greater interaction between the external flow and the cavity flow through enhanced shear layer mixing, thereby promoting flow penetration into the cavity. Irrespective of the $L/D$ ratio and $Re_D$, the $St$ at the cavity downstream location (see Figure \ref{fig:HF_LbyD_effect_0p5t2} (c)) is observed to be higher than that at the cavity rear face location (see Figure \ref{fig:HF_LbyD_effect_0p5t2}(b)), while remaining lower than the incoming flow $St$ (see Figure \ref{fig:HF_LbyD_effect_0p5t2}(a). Furthermore, for $L/D = [0.5, 1.0, 2.0]$, the laminar shear layer impacts on the rear wall, resulting in a reattachment shock. At higher Reynolds numbers, the flow recovers due to stronger reattachment shock, and the Stanton number approaches that of the incoming flow and $St$ at $(S_T)_d$ remains nearly constant for most $L/D$ values when ($Re_D \geq 45200$).

\begin{figure}[hbt!]
\centering
	\includegraphics[width=0.7\textwidth]{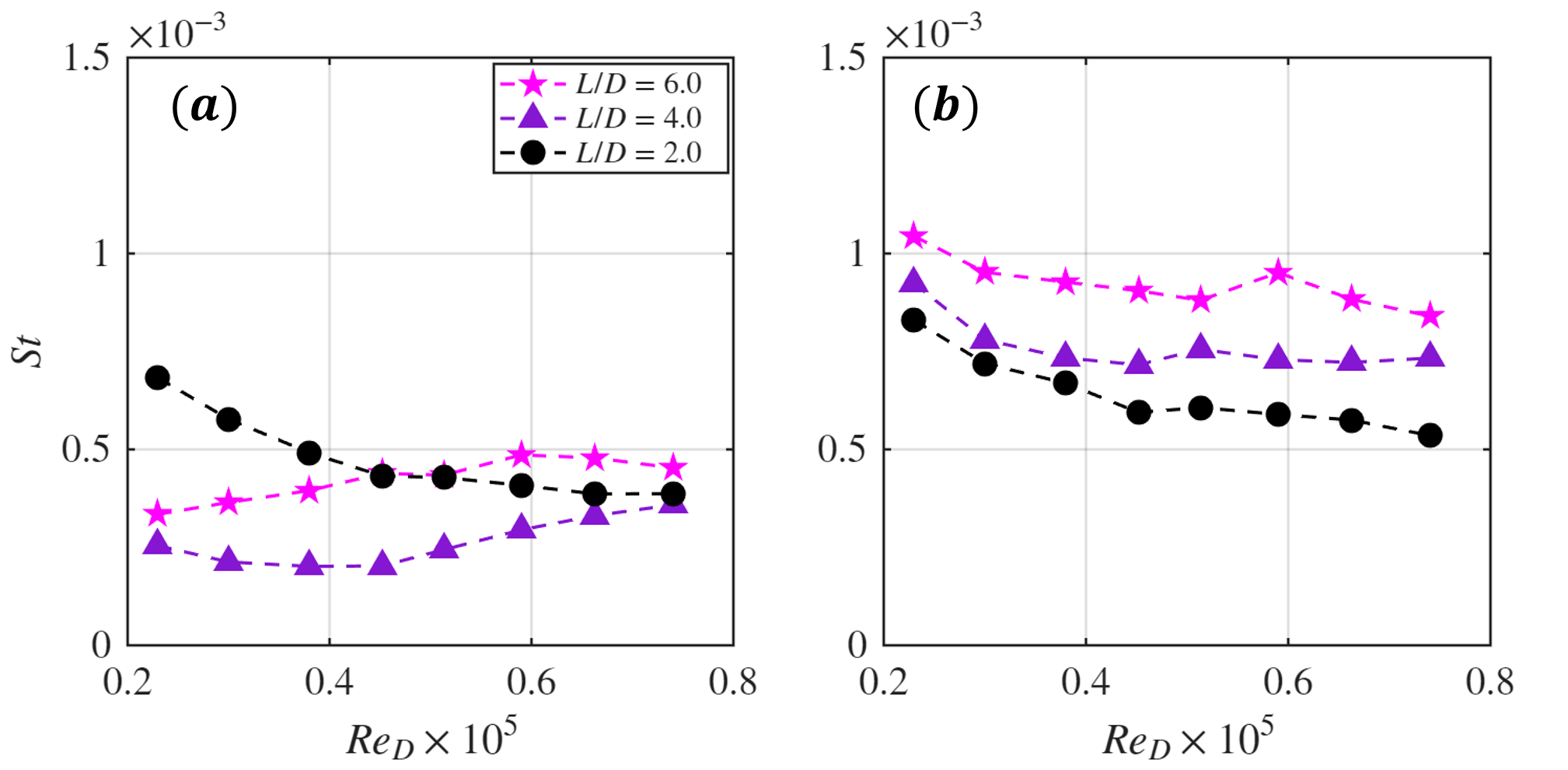}
	\caption{Local Stanton number variation obtained from (a) $(S_T)_{r}$  (b)  $(S_T)_{d}$  (See Figure \ref{fig:Sensor} for sensor details) over $23000\leq Re_D \leq 74000$ for different $L/D$ ratios of 2.0 , 4.0 and 6.0. Sensor $(S_T)_{u}$ is unavailable for $L/D = 4.0, 6.0$ due to space constraints.}
    \label{fig:HF_LbyD_effect_2t6}
\end{figure}

Figure \ref{fig:HF_LbyD_effect_2t6} presents the Stanton number measured on the cavity rear wall ($(S_T)_r$) and the downstream surface ($(S_T)_d$) for $L/D = [2.0, 4.0, 6.0]$ over the range of Reynolds numbers investigated. Due to space constraints, the upstream cavity sensor $((S_T)_u)$ data are not presented for $L/D = 4.0$ and 6.0. For $L/D = 4.0$, the Stanton number on the cavity rear wall (Figure \ref{fig:HF_LbyD_effect_2t6}(a)) is significantly lower than that for $L/D = 2.0$ at relatively low $Re_D$, where the separated shear layer remains laminar. For $L/D=4.0$, at lower $Re_D$ shear layer is predominately laminar and as $Re_D$ increases to $45200$ the formation of K-H vortices can be observed (see Figure \ref{fig:Sch_lbyd}-IV). This phenomena is reflected in $St$ variation at $(S_T)_r$ which remains almost constant for initial $Re_D$ cases, and linear increment is seen for $Re_D \geq 45200$. A similar trend is observed for $L/D = 6.0$. However, owing to the longer cavity length, formation of the K-H vortices in the shear layer appears at lower $Re_D$s, which eventually results in higher $St$ in comparison to $L/D=4.0$. At higher $Re_D$s the turbulent transitioning of the shear layer substantially enhances convective heat transfer, yielding higher rear-wall Stanton numbers than those observed for the other $L/D$ ratios.

Irrespective of the $L/D$ and $Re_D$, the Stanton number in the cavity downstream location (see Figure \ref{fig:HF_LbyD_effect_2t6}(b)) is found to be higher in comparison to the cavity rear face location (see Figure \ref{fig:HF_LbyD_effect_2t6} (a)).This increment is most likely due to the reattachment shock at the rear wall of the cavity owing to shear layer impingement. The local $St$ at the downstream cavity location is higher for higher $L/D$ (see Figure \ref{fig:HF_LbyD_effect_2t6}(b)), whereas the laminar reattachment results in a lower $St$ for the cases of $L/D = 2.0$. Also, $(S_T)_{d}$ for the $L/D=2.0$ and $4.0$ remains almost constant for $Re_D \geq 38000$. Nevertheless, for $L/D = 6.0$, the slope of $St$ seems to change for $Re_D\geq 59000$ for both $(S_T)_r$ and $(S_T)_d$ locations, which can be attributed to the increase in the local heat flux due to the transition of the shear layer to a turbulent state\cite{nanda2026}. Moreover, the transitional $Re_D$ also matches the PLRS-based FFT analysis for the dominant mode reported in our previous work \cite{nanda2026}, as observed in Figure \ref{fig:HF_LbyD_effect_2t6}. It can be noticed that, for the case of $L/D = 6.0$, at higher $Re_D$s, the $St$ corresponding to the $(S_T)_{d}$ is more than that of the $(S_T)_{u}$ emphasizing the fact of having higher heat flux for the turbulent flows as compared with the attached flows.

\subsection {Rear Wall Height Effect}

\begin{figure}[hbt!]
\centering
\includegraphics[width=0.8\textwidth]{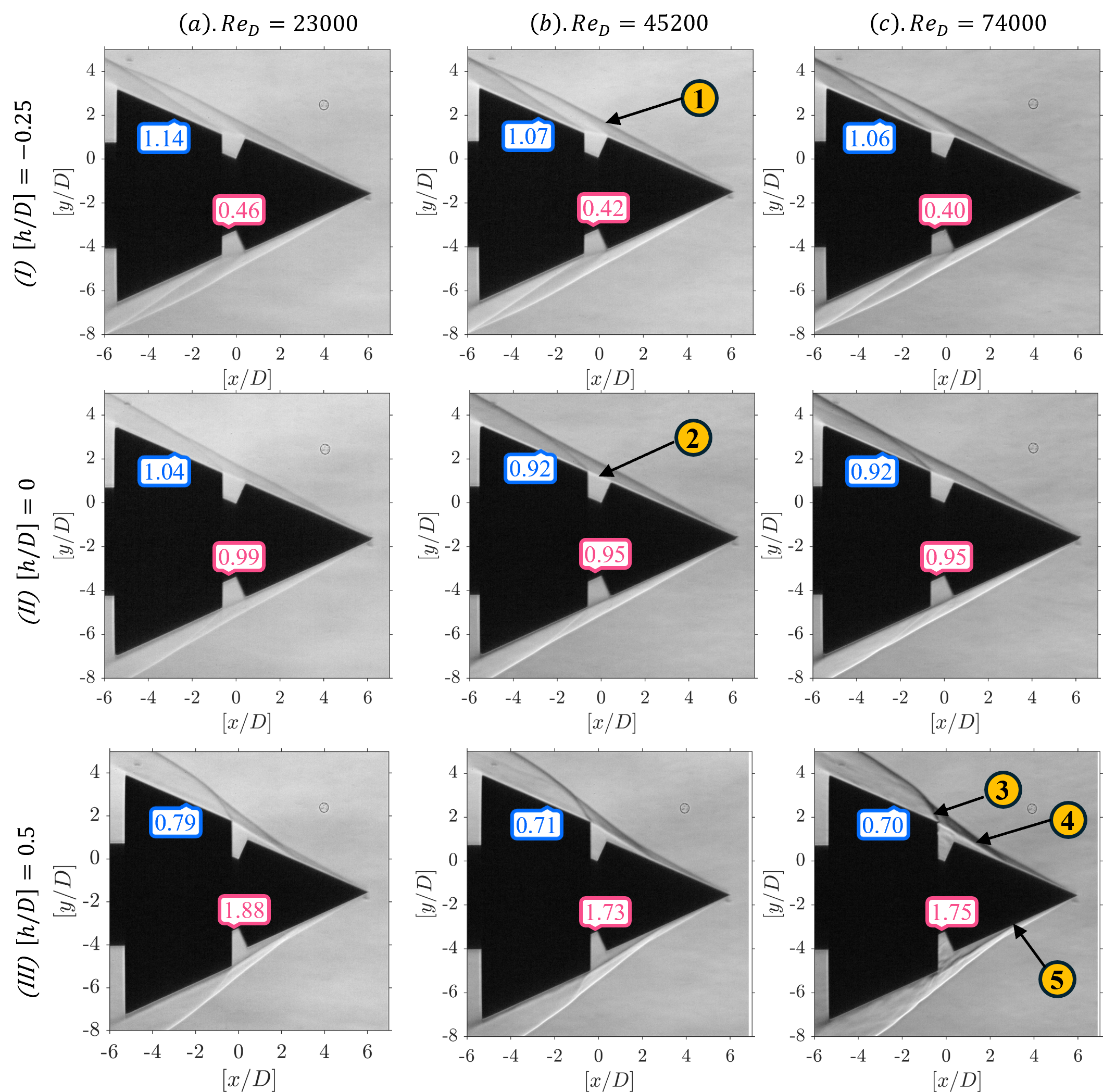}
\caption{Instantaneous schlieren images for $L/D=1.0$ and $[h/D] = [-0.25,0,0.5]$. Flow is from right to left at $Re_D = [2300,45200,74000]$. Key flow features: 1. oblique shock from the cone tip, 2. separated shear layer, 3. reattachment shock, 4. separation shock, 5.upstream flow separation point. }  
\label{fig:Schliren_hbyD_1}
\end{figure}

\subsubsection{ Schlieren Flow Visualization }

\begin{figure}[hbt!]
\centering
\includegraphics[width=0.8\textwidth]{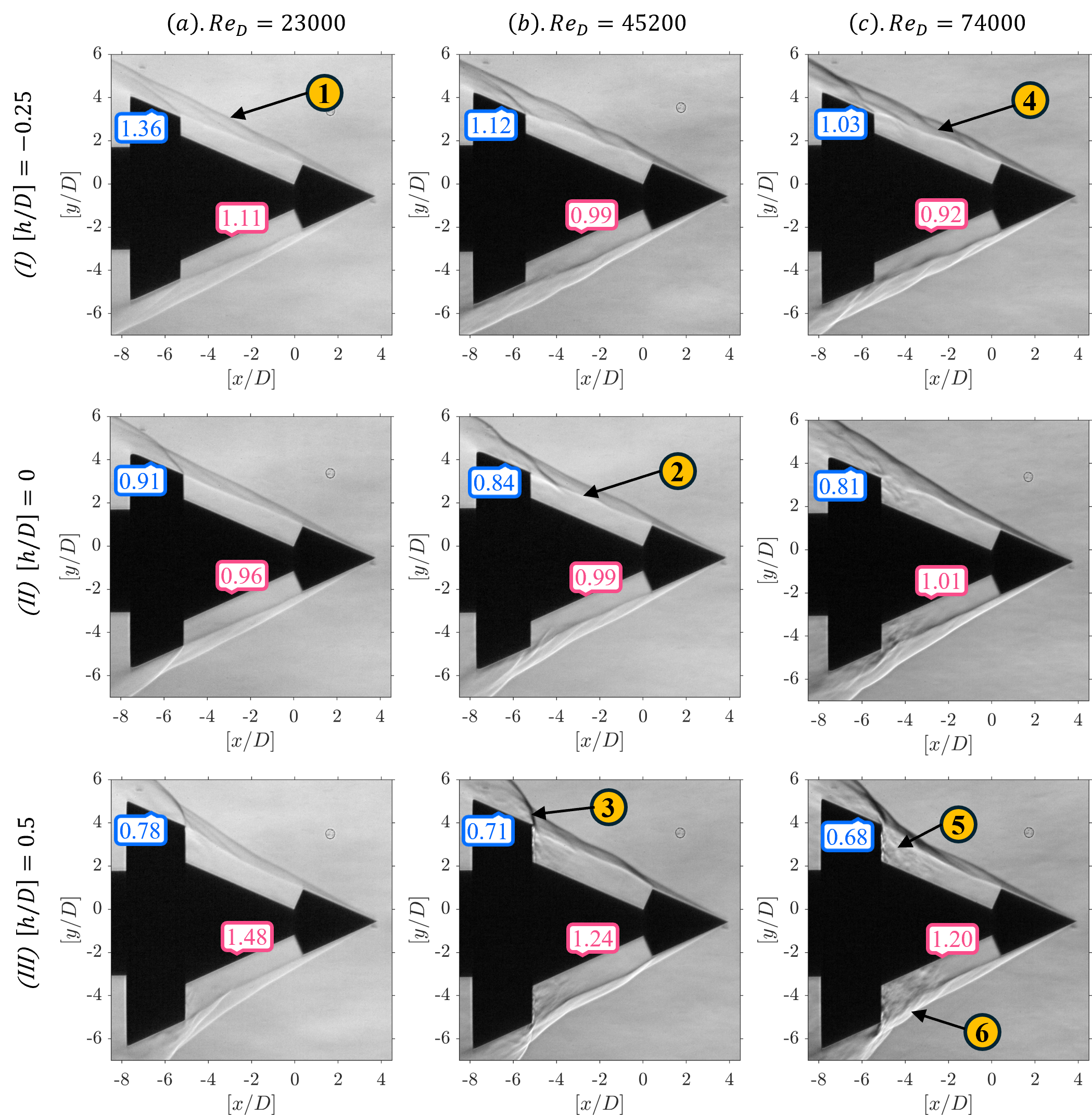}
\caption{Instantaneous schlieren images for $L/D=6.0$, $[h/D] = [-0.25,0,0.5]$. Flow is from right to left at $Re_D = [2300,45200,74000]$. Key flow features:  1. leading edge conical shock, 2. separated shear layer, 3. reattachment shock at the trailing edge, 4. shock wave from the undulating shear layer, 5. upstream propagation of acoustic waves, and 6. shear layer turbulent breakdown. }
\label{fig:Schliren_hbyD_6}
\end{figure}

The effects of rear-wall height-to-depth ratio, $h/D=-0.25, 0$ and $0.5$ for $L/D=1.0$ at $Re_D=23000,45200$ and 74000 are shown in Figure \ref{fig:Schliren_hbyD_1}. For the negative trailing-edge configuration of $h/D=-0.25$ (Figure \ref{fig:Schliren_hbyD_1} -I), increasing $Re_D$ strengthens the reattachment shock, as the flow impinged downstream of the trailing edge, resulting in higher pressure at $(S_p)_d$ location relative to the other case of $h/D$. As observed previously, higher ($Re_D$) also contributes to the downstream pressure-recovery. The comparatively lower cavity floor pressure for $h/D=-0.25$ (see Figure \ref{fig:Schliren_hbyD_1} -I) further indicates that the recirculation region is relatively isolated from the outer flow as the shear layer reattaches downstream.
In contrast, the positive trailing-edge configuration ($h/D=0.5$) induces upstream flow separation, as shown in Figure \ref{fig:Schliren_hbyD_1}-III, causing the flow to leave the cavity rear edge tangentially. This behavior persists across all Reynolds numbers, and with increasing $Re_D$, the separation shock gets strengthened along with upstream shifting of the onset of separation. As the incoming flow is exposed to the separation shock for $h/D=0.5$, thereby results in significant increment in the pressure at the $(S_p)_{fc}$ location. Also, the reattachment shock gets steeper in comparison to other $h/D$ cases (compare Figure \ref{fig:Schliren_hbyD_1}-I and Figure \ref{fig:Schliren_hbyD_1}-III), effectively results in reduced downstream pressure recovery, which is also observed during the aspect ratio variations.

Figure \ref{fig:Schliren_hbyD_6} shows schlieren visualizations of the effect of rear-wall height-to-depth ratio ($h/D$) for a cavity with $L/D=6$ at $Re_D=23000$, $45200$, and $74000$. For $h/D=-0.25$, owing to the lesser rear wall height in conjunction with intermittency in K-H structures shedding in the shear layer, the flow at times reattaches downstream, leading to increased $\zeta$ at $(S_p)_d$ (see Figure \ref{fig:Schliren_hbyD_6} -I). Consistent with the $L/D=1$ observations in the earlier discussion, the reattachment shock steepens with increasing $h/D$, resulting in a smaller $\zeta$ for higher $h/D$ cases. As $h/D$ increases to $0.5$, the pressure build-up inside the cavity \citep{nanda2026} drives out-of-plane motion of the entire separated shear layer and associated shock system, resulting in a significant increase in $\zeta$ at $(S_p)_{fc}$. The resulting strong oscillatory motion seems to bear similar characteristics to a flapping shock–shear-layer interaction reported in hypersonic leading-edge separated flows \cite{Karthick_2023} and supersonic spike-body configurations \cite{Feszty,sahoo2021shock}.

\subsubsection{ Surface Heat Flux Measurement }

\begin{figure}[hbt!]
\centering
	\includegraphics[width=1.0\textwidth]{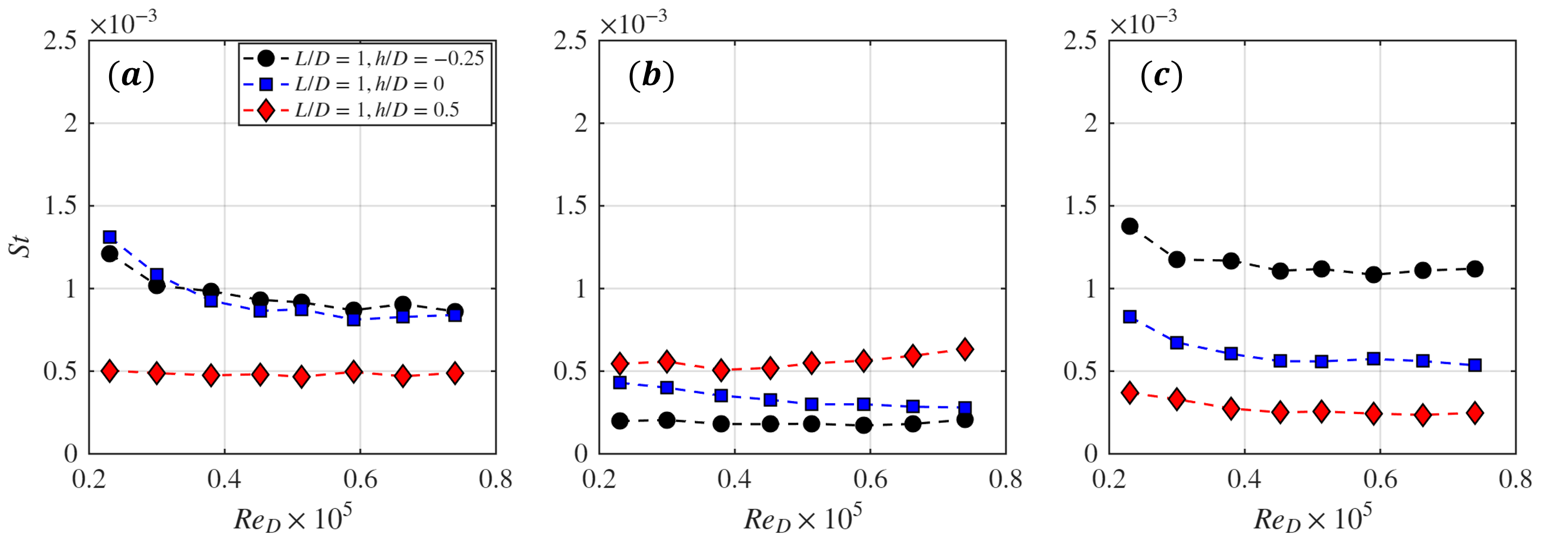}
	\caption{Local Stanton number variation obtained from (a)$(S_T)_{u}$ (b) $(S_T)_{r}$ (c)  $(S_T)_{d}$  (See Figure \ref{fig:Sensor} for sensor details) over $23000\leq Re_D \leq 74000$ for different $h/D$ for $L/D=1.0$.}
    \label{fig:HF_HbyD_Lbyd_1_effect}
\end{figure}

The local Stanton number is determined from the temperature rise measured by the CSJT at the three locations shown in Figure \ref{fig:Sensor} and the variation of the Stanton number with $Re_D$ and $h/D$ for $L/D = 1.0$ is shown in Figure \ref{fig:HF_HbyD_Lbyd_1_effect}. At the upstream location ($(S_T)_u$), the $St$ values for $h/D = -0.25$ and $0$ exhibit nearly identical trends with variation of $Re_D$, as the flow remains attached to the cone surface in both cases (see Figure\ref{fig:Schliren_hbyD_1} I-II and Figure \ref{fig:HF_HbyD_Lbyd_1_effect} (a)). In contrast, the $St$ values for $h/D = 0.5$ are substantially lower than those for the other two cases, as shown in Figure \ref{fig:HF_HbyD_Lbyd_1_effect}(a). This behavior is mainly attributed to flow separation well upstream of the cavity leading edge, owing to the satisfy the condition of flow leaving the cone tangentially (Figure \ref{fig:Schliren_hbyD_1}-III). The $St$ at the cavity rear-wall location is presented in Figure \ref{fig:HF_HbyD_Lbyd_1_effect} (b). For $h/D=-0.25$ and $0$, the interaction point between the oblique shock originating from the cone tip and the reattachment shock shifts downstream (Figure \ref{fig:Schliren_hbyD_1} I-II). This downstream interaction owing to the relatively swallower reattachment shock angle reduces the intensity of the flow disturbances reaching the cavity rear wall. This observation is further supported by the pressure measurements, which indicate that the recirculation bubble remains strongly isolated from the outer cavity flow, resulting in relatively lower $St$ values. In contrast, for $h/D=0.5$, the oblique shock from the cone tip interacts with the separation shock in the vicinity of the cavity rear face, as observed in the instantaneous schlieren image (Figure  \ref{fig:Schliren_hbyD_1}-III). This interaction aids in enhancing local heat transfer inside the cavity. Consequently, the $St$ at the cavity rear-wall location is significantly higher for $h/D=0.5$ than for the other cases (Figure \ref{fig:HF_HbyD_Lbyd_1_effect}(b)). A similar analogy of flow reattachment downstream of the cavity (Figure \ref{fig:Schliren_hbyD_1}-I) leads to the significantly higher value of $St$ at the $(S_T)_d$ location for $h/D=-0.25$, as shown in Figure \ref{fig:HF_HbyD_Lbyd_1_effect} (c). With increasing $h/D$, the $St$ at the $(S_T)_d$ location reduces irrespective of the $Re_D$. Also, $St$ remains nearly constant at higher $Re_D$s for all the sensor locations and $h/D$ values. 

\begin{figure}[hbt!]
\centering
	\includegraphics[width=0.65\textwidth]{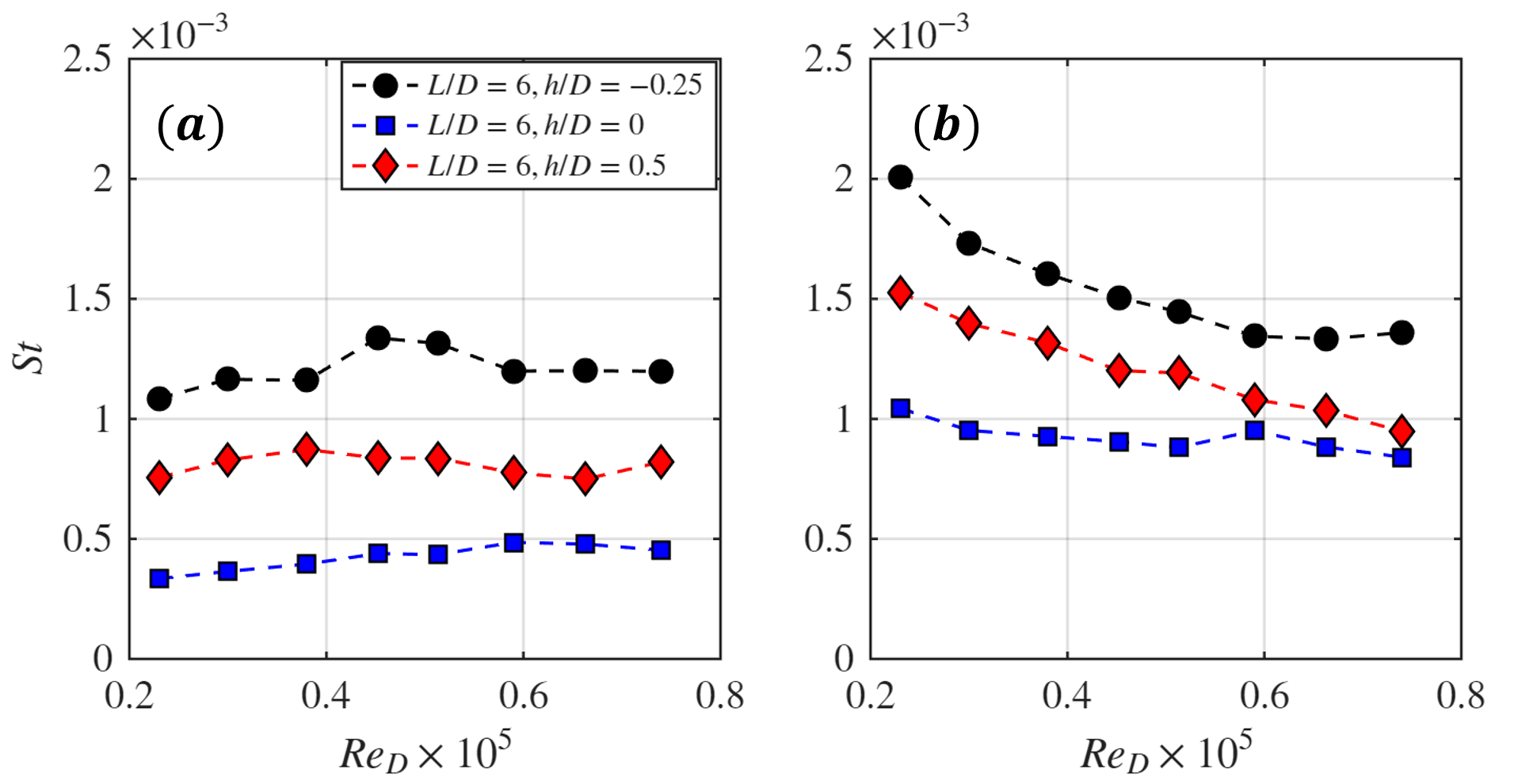}
	\caption{Local Stanton number variation obtained from (a)$(S_T)_{r}$ (b) $(S_T)_{d}$   (See Figure \ref{fig:Sensor} for sensor details) over $23000\leq Re_D \leq 74000$ for different $h/D$ for $L/D=6.0$.}
    \label{fig:HF_HbyD_Lbyd_6_effect}
\end{figure}

The local Stanton number variation obtained on the cavity downstream and rear face at different $h/D$ for $L/D=6.0$ is shown in Figure \ref{fig:HF_HbyD_Lbyd_6_effect}. The $St$ trend variation for $h/D = 0$ has already been discussed in the previous section. As expected, the local Stanton number near the reattachment location, that is, at the $(S_T)_{d}$ location, is greater than that of the $(S_T)_{r}$ irrespective of the $h/D$. As noted during previous discussions, owing to the intermittency of the K-H structure imparting downstream and at the rear wall for the $h/D=-0.25$ case at different time instances, $St$ is found to be maximum among the cases for both the $(S_T)_{r}$ and $(S_T)_{d}$ location. Also, the flapping of the entire shock-shear layer system resulting in continuous in-flux and ejection of the flow for $h/D=0.5$ seems to aid in the heat transfer in comparison to $h/D=0$ case. Due to space constraints, the $(S_T)_{u}$ sensor is absent for these cases. However, considering the invariant nature of the incoming flow with $L/D$, upon comparison with the earlier obtained $St$ variation (see Figure \ref{fig:HF_HbyD_Lbyd_6_effect} (c)), it is found that irrespective of the positive or negative $h/D$, the $(S_T)_{d}$ location have higher $St$ in comparison to the incoming flow. 

\section{Conclusion}
  Experiments were conducted on a hypersonic axisymmetric cavity model to quantify the effects of shock and shear-layer unsteadiness on $St$ for different cavity configurations. Experiments are conducted in the Hypersonic Ludwig Tunnel (HLT) at Technion with a freestream Mach number of $M_\infty=6.0$. Different flow and geometrical conditions are studied. Reynolds number based on cavity depth ($Re_D$) is varied between 23000 to 74000. The cavity length-to-depth ratio ($L/D$) is varied between 0.5 and 6.0. In addition, the excess height-to-depth ratio ($h/D$) of the cavity's trailing edge is also varied between -0.25 to 0.5. The CSJTs are mounted upstream, inside and downstream of the cavity in order to assess their variations. The following are the major conclusions from the present study.
  
\begin{itemize}
 \item{\textbf{Effect of $L/D$:} As the cavity aspect ratio $(L/D)$ increases, the downstream Stanton number $(St_d)$ increases. When the shear layer undergoes transition to turbulence, it strengthens mixing and enhances convective heat transfer downstream of the cavity. As a result, the downstream $St_d$ becomes higher than the Stanton number associated with the incoming attached flow.}
 
\item{\textbf{Effect of $[h/D]$:} For the positive rear wall configuration $(h/D=0.5)$ at $L/D=1$, upstream shear-layer separation results in the lowest downstream cavity Stanton number $(St)$. In contrast, for the negative excess rear wall configuration $(h/D=-0.25)$ at $L/D=1$ and $6$, downstream reattachment substantially enhances the local heat flux downstream of the cavity, yielding values that exceed those of the incoming attached flow.}
\end{itemize}

\section*{Acknowledgments}
The authors gratefully acknowledge the invaluable contributions of lab engineers including M. Dunaevsky, O. Kan, N. Shefer, D. Naftali, and E. Shulman, for their significant assistance in conducting these experiments.

\bibliography{sample}

\end{document}